\documentclass[12pt]{iopart}

\usepackage{array}
\usepackage{booktabs}
\usepackage{multirow}

\usepackage{iopams} 
\usepackage{graphicx}
\usepackage{cite}
\usepackage{color}
\usepackage{soul}

\begin{document}

\title[Energy efficiency of He metastable generation via VWT in $\mu$APPJs]{Energy efficiency of voltage waveform tailoring for the generation of excited species in RF plasma jets operated in He/N$_2$ mixtures}

\author{I Korolov$^1$, Z Donk\'{o}$^2$,  G H\"{u}bner$^1$, Y Liu$^1$, T Mussenbrock$^1$, J Schulze$^{1,3}$}

\address{$^1$Department of Electrical Engineering and Information Science, Ruhr-University Bochum, D-44780,
Bochum, Germany}

\address{$^2$ Wigner Research Centre for Physics,
H-1121 Budapest, Konkoly-Thege Mikl\'os str. 29-33, Hungary}

\address {$^3 $Key Laboratory of Materials Modification by Laser, Ion and Electron Beams, School of Physics, Dalian University of Technology, China}

\ead{korolov@aept.rub.de}

\begin{abstract}

Based on Tunable Diode Laser Absorption Spectroscopy (TDLAS) measurements of the spatially averaged and peak helium metastable atom densities in a capacitively coupled micro atmospheric pressure plasma jet operated in He/N$_2$ mixtures, the energy efficiency of metastable species (He-I 2$^3$S$_1$) generation is compared for three different scenarios: single frequency operation at (i) 13.56 MHz and (ii) 54.12 MHz, and Voltage Waveform Tailoring (VWT) at (iii) ``valleys"-waveforms synthesized from four consecutive harmonics of 13.56 MHz. For each case, the dissipated power is measured based on a careful calibration procedure of voltage and current measurements. The results are compared to PIC/MCC simulation results and very good agreement is found. The computational results show that the choice of the surface coefficients in the simulation is important to reproduce the experimental data correctly. Due to the enhanced control of the spatio-temporal electron power absorption dynamics and, thus, of the electron energy distribution function by VWT, this approach does not only provide better control of the generation of excited and reactive species compared to single frequency excitation, but in case of helium metastables the energy efficiency is also shown to be significantly higher in case of VWT. 

\end{abstract}

\section{Introduction}

Non-thermal atmospheric pressure microplasmas have received  rapidly  growing  attention from academic as well as from applied fields. The reactive species generated in such plasmas are important for applications  such  as  plasma medicine (wound healing, cancer treatment, etc.) \cite{Bekeschus2016, Boeckmann2020, Xu2015, Schmidt2017, Hui2020, Binenbaum2017, Haertel2014, Bekeschus2017, Nicol2020}, surface treatment/modification\cite{Pawlat2016, Mozetic2010, Saw2016, Chou2017, Dey2020, Sener2020},  plasma catalysis \cite{Abiev2020, Yayci2020, Neyts2015}, nanoparticle and nanocrystal synthesis\cite{Haq2019, Mariotti2016,Kortshagen2016, Lazea-Stoyanova2015}, etching/deposition \cite{Benedikt2007,Ichiki2004,Reuter2012p}, etc. The main reason for the popularity of such plasma sources is their ability to generate reactive particle species including, e.g., atomic oxygen and/or nitrogen species (RONS), at temperatures close to ambient air temperature. Microscopic atmospheric pressure plasma jets ($\mu$APPJ) are one of the widely used instruments developed and used to generate such plasmas. They are typically operated at a single frequency using argon or helium as a buffer gas with a small admixture of molecular gases such as oxygen, nitrogen, hydrogen and water vapour ~\cite{Schroter2020, Stoffels2006, Gorbanev2019, Volkov2018, Reuter2008, Benedikt2018, Benedikt2016, Ellerweg2012, Schneider2014, Yue2020, Klose2020, Kondeti2020}. 

\noindent
Recent experimental and computational studies have shown that the generation of RONS can be further optimized and controlled by using customized driving voltage waveforms \cite{Korolov2019, Gibson2019, Korolov2020, Korolov2021}. In general, this method of `voltage waveform tailoring' (VWT) is based on generating customized voltage waveforms based on Fourier synthesis from $N$ consecutive harmonics of a fundamental frequency, $f_0$. In contrast to single-frequency atmospheric pressure discharges, where the electron power absorption dynamics is inherently symmetric in space between the electrodes and in time within one radio-frequency (rf) period, VWT allows to break and control the spatio-temporal symmetry of the electron power absorption dynamics. By using, e.g., "peaks"- or "valleys"-waveforms, electron power absorption can be confined into a small region in space and to specific times within the fundamental RF period. As a consequence, most of the power is dissipated to a group of electrons within this small spatio-temporal region of interest, whose location and size can be adjusted by tuning the driving voltage waveform. Such confinement allows one to control the high energy tail of the electron energy distribution function (EEDF) in this region and, thus, the generation of excited species and radicals \cite{Korolov2019, Liu2021VWT}. An excited species of particular interest is helium metastables, which are an important source of stored energy and play a key role for plasma generation via Penning ionization in $\mu$APPJs operated in Penning mixtures, e.g, He/N$_2$, He/O$_2$ \cite{Bischoff2018, Waskoenig2010, Zhang2006, Spiekermeier2015, Liu2010}. VWT has previously been demonstrated to provide control of their density, to optimize plasma generation and reactive particle densities \cite{Korolov2020, Korolov2021}.

\noindent
Recent studies of VWT in $\mu$APPJs were mainly focused on comparisons of the densities of relevant species with those generated in single frequency discharges operated at identical peak-to-peak driving voltages \cite{Gibson2019, Korolov2021, Liu2021VWT}.  However, two important fundamental questions, that are highly relevant for applications, remained open: (i) How does the energy efficiency of generating selected species via VWT compare to using single driving frequencies? (ii) How much energy and at what spatial location is dissipated in such systems? - The key to successfully answering these questions is to determine correctly the real power absorbed by the plasma.  Frequently in the literature, only the generator power is mentioned \cite{Knake2008, Bibinov2011, Niemi2011, Waskoenig2010, Zhu2005}. Moreover, prior measurements and theoretical investigations of the electrical power dissipated in atmospheric pressure plasma jets have mainly been restricted to the single-frequency case \cite{Reuter2018, Gerling2017, Golda2016, Golda2019, Golda2020, Dunnbier2015, Chirokov2009}. Here, we present measurements of the dissipated power and metastable densities in $\mu$APPJs driven by different single frequencies  and tailored voltage waveforms in He/N$_2$ mixtures. The experimental findings are compared to the results obtained from Particle-in-Cell simulations with Monte Carlo treatment of collisions (PIC/MCC). Our findings show that VWT allows to generate helium metastables in a more energy efficient way compared to single driving frequencies. Moreover, it allows to control the densities of selected particle species that are relevant for distinct applications, more easily as compared to classical single frequency $\mu$-APPJs.

\noindent
The paper is structured in the following way: in the second section, the experimental set-up, the method to obtain the power dissipated in the $\mu$APPJ and to measure the helium metastables density, as well as the simulation approach are described briefly. In the next section, the results are presented and discussed. In this section, the effects of the power absorbed by the jet and the generation of helium metastables are studied for three different scenarios: single frequencies at (i) 13.56 MHz and (ii) 54.12 MHz, as well as (iii) "valleys"-waveforms constructed from $N = 4$ consecutive harmonics of the fundamental frequency $f_0$ =  13.56 MHz.  Finally, conclusions are drawn in section 4.   

\section{Experimental set-up and PIC/MCC simulations}

\subsection{Experimental set up}

\begin{figure}[!htb]
\begin{center}
\includegraphics[width=0.8\textwidth]{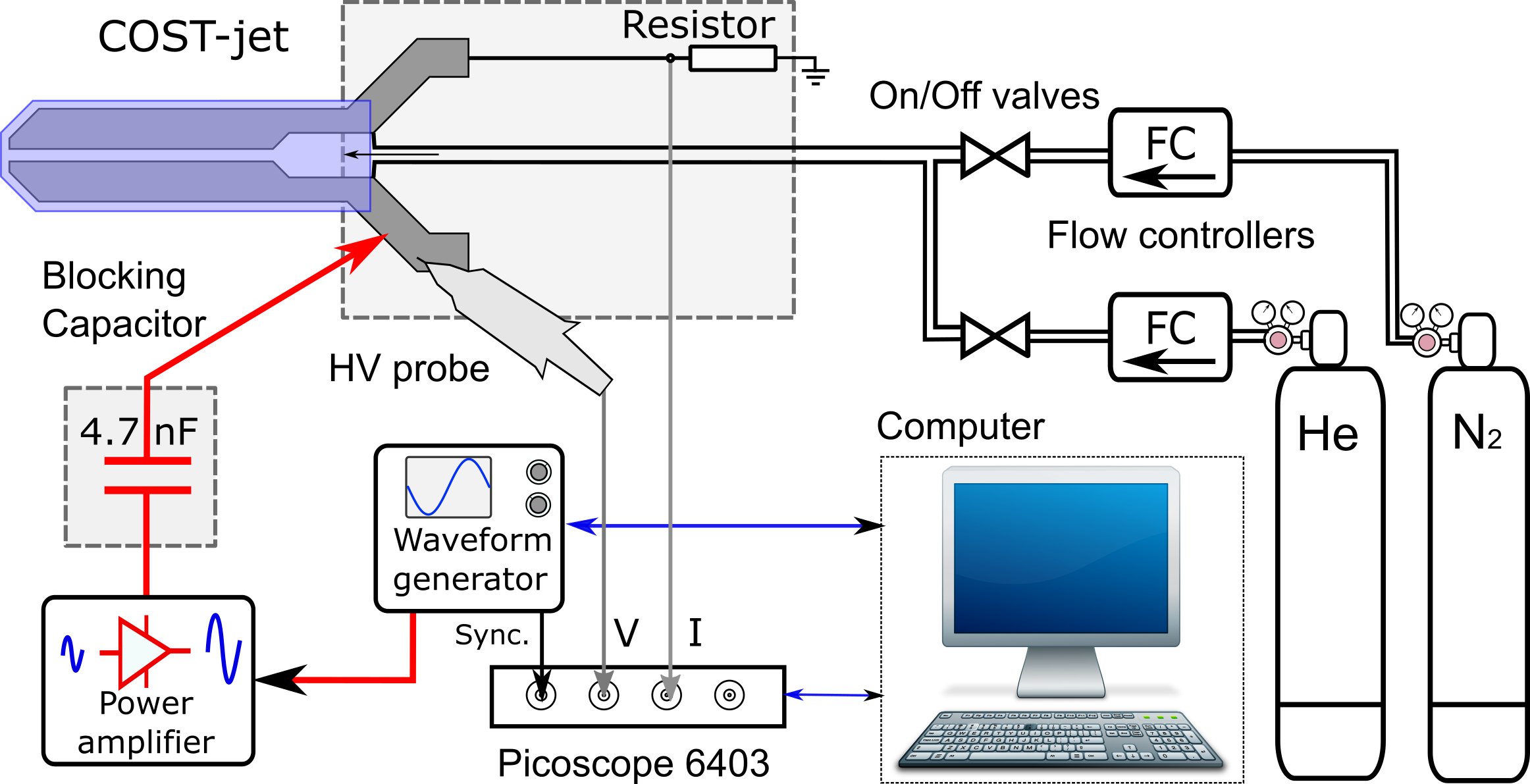}
\end{center}
\caption{Schematic view of the experimental set-up}
\label{fig:setup}
\end{figure}

\noindent
Figure~\ref{fig:setup} shows a sketch of the experimental set-up. The measurements are performed using a reference microplasma ``COST-jet" developed within the framework of the European Cooperation in Science and Technology. The plasma volume of the jet of $30\times1\times1$ mm$^3$ is confined between two stainless steel electrodes and two quartz plates. The electrode gap is fixed at 1 mm. More detailed information about the plasma source can be found elsewhere \cite{Golda2016}. The discharge is operated in helium (5.0 purity) with 0.1 \% nitrogen (5.0 purity) admixture. The gas handling system has been described in \cite{Bischoff2018, Korolov2019}. The flow rates are fixed at 1000 sccm for helium and at 1 sccm for nitrogen. The gas pressure inside the COST-jet is 1.02($\pm0.02$)$\times 10^5$ Pa.

\subsubsection{Voltage Waveform Tailoring\\}

\noindent
The COST-jet is typically operated with a single driving frequency of 13.56 MHz for which an internal coupling circuit is used for impedance matching \cite{Golda2016}. In order to generate a plasma by tailored voltage waveforms, the internal matching circuit was removed. The new connection scheme is shown in figure~\ref{fig:setup}. The desired waveform is generated using an arbitrary waveform generator (Keysight 33600A). The signal is then applied to the powered electrode through a broadband power amplifier (Vectawave VBA250-400) and a blocking capacitor (4.7 nF). The voltage waveform at the electrode is measured using a high voltage probe (Tektronix P6015A, 75 MHz) and an USB oscilloscope (Picoscope 6402C, 250 MHz bandwidth, 5 Gs/s). The signal is examined and waveform correction factors are calculated using a LabVIEW software. A corrected signal is then  sent to the waveform generator. This corrective feedback loop is repeatedly executed until the deviation between the measured and desired phases/amplitudes of all harmonics, used to create the waveform, reaches values of 1 - 3 \% at the electrode. Since the power is directly applied to the jet without any matching, $>$95-99\% of the applied power is reflected. For our fundamental studies the reflected power is not an issue. For applications, impedance matching in the presence of such multi-frequency tailored voltage waveforms can be realized based on recently developed multi-frequency impedance matching networks \cite{Schmidt2018,Match1,Match3}.

\noindent
The driving voltage waveforms can be expressed using the following equation: 

\begin{equation}
\phi(t) =\sum_{k=1}^{N}\phi_k \cos(2\pi k f_0 t + \theta_k),
\label{eq:voltage}
\end{equation}

\noindent
where $N$ is the number of consecutive harmonics of the fundamental frequency $f_0$ and $\phi_k$ are the amplitudes of the individual harmonics. The phase angles, $\theta_k$, are set to $0^\circ$ for all harmonics for "peaks" waveforms, while the phases of the even harmonics are set to 180$^\circ$ for "valleys" waveforms. The voltage amplitudes of the harmonics are calculated according to: $\phi_k=\phi_0(N-k+1)/N$, where $\phi_0= 2 \phi_{\rm pp} N/(N+1)^2$  \cite{Schulze2011eae}. $\phi_{\rm pp}$ is the peak-to-peak value of the driving voltage waveform (when peaks- or valleys-waveforms are set.) The maximum of $\phi_{\rm pp}$ is limited by the requirement to generate a stable discharge, i.e. any kind of arcing/constricted mode at high voltages must be avoided. The comparison of the energy efficiency of the generation of helium metastable atoms is carried out for the following cases: (1) single frequency ($N$ = 1) at $f_0$ = 13.56 MHz, (2) single frequency ($N$ = 1) at $f_0$ = 54.24 MHz and (3) "valleys" waveforms at $f_0$ = 13.56 MHz and $N$ = 4. In this way, results obtained for a tailored voltage waveform are compared to those obtained in single frequency plasmas generated with the lowest and highest frequency component used to construct the customized driving voltage waveform.

\subsubsection{Power measurements\\}

\noindent
A method to determine the power dissipated in a single frequency COST-jet driven at 13.56 MHz is described by Golda et.al in \cite{Golda2016}. This method is based on the relation: 
\begin{equation}
P = 0.5 I U \cos(\alpha),
\label{eq:psingle}
\end{equation}
 where $U$, $I$ and $\alpha$ is the voltage and current amplitude, as well as the phase shift between the current and voltage waveform, respectively. The current is calculated from the voltage drop across an internal resistor (4.7 $\Omega$, see figure~\ref{fig:setup}) included in the COST-jet. In order to minimize the noise level, the internal input resistance of the oscilloscope is set to 50 $\Omega$. The voltage drop does not exceed a few volts for the highest powers used in this paper. Without plasma, the jet represents a capacitor and, thus, the phase shift between voltage and current is expected to be $\alpha = 90^\circ$. Due to the voltage probe response, time delays in the electronics, cables, etc., the measured phase angle without plasma can be remarkably different and a calibration/correction must be performed in the absence of the discharge. 
  \noindent
 The situation becomes more complex, when a higher number of harmonics is applied ($N > 1$). In general, for periodic voltage and current signals with a period $T$, the power can be expressed as follows: 
  \begin{equation}
P_{\rm{tot}} = \frac{1}{T} \int_{0}^{T} I(t)U(t)\, dt. 
\label{eq:ptotint}
\end{equation}
 
 \noindent
 The time dependent current, $I(t)$, and voltage, $U(t)$, waveforms now contain multiple harmonics. Both waveforms can be represented as a Fourier series. One way to determine the dissipated power is to perform the calibration procedure described for a single frequency discharge for each Fourier component separately in the absence of the plasma. Based on this, the current and voltage waveforms measured in the presence of a discharge can then be Fourier transformed and each Fourier term can be analyzed separately. Then the current and voltage waveforms can be re-constructed (including the calibration) and substituted into eq.~(\ref{eq:ptotint}). If other non-harmonic frequency components were self-excited by the plasma, this analysis would be quite challenging, since a variety of phase shifts between all frequency components would have to be obtained accurately. This is, for instance, frequently the case in low pressure capacitive RF plasmas \cite{Schuengel2015PSR}, but not in $\mu$APPJs (as verified by our measurements), where voltage and current contain only harmonics of the base frequency, $f_0$. In this case, we get \cite{Emanuel1990}:
 
\begin{equation}
P_{\rm{tot}} = I_{\rm{dc}}U_{\rm{dc}} +0.5\sum_{\rm{k}=1}^{N} I_{\rm{k}}U_{\rm{k}}\cos(\alpha_{\rm{k}}).
\label{eq:ptot}
\end{equation}

\noindent
The first term takes into account possible power dissipation due to  dc self bias voltage generation, $U_{\rm{dc}}$, for $N > 1$, see \cite{Korolov2019}. For the present conditions, this term can be neglected, since it is  within the noise level of the measurements and $<2\%$ of the total power. $U_{\rm{k}}$, $I_{\rm{k}}$ and $\alpha_{\rm{k}}$ is the voltage and current amplitude of the $k$-th harmonic as well as the phase shift between voltage and current of each harmonic, respectively. 

\noindent
In this work, the current and voltage waveform measurements are triggered by the waveform generator in order to minimize the sampling jitter of the signals and, thus, to increase the precision of the power calculation.  The signals are recorded by a digital oscilloscope (Picoscope 6402C) with 0.8 ns step resolution. The triggered signals with a length of 10 000 samples are stored and subsequently averaged for 1000 cycles. Calibration/recording/correction of the signals and the determination of the power values is performed by a LabView software. Using the same COST-jet, the dissipated power values measured on different days  did not differ by more than 2-5\%, confirming the stability of the system. The estimated error of the measurements is 10-15\%.

\subsubsection{Helium metastable density measurements\\}

The helium metastable density in the plasma is  measured by Tunable Diode Laser Absorption Spectroscopy (TDLAS). To obtain absolute densities, the absorption profile of two He-I triplet transitions ($2^3$S$_1$ $\rightarrow$ $2^3$P$_{{J}}$, where $J$ = 1,2) with central wavelengths of $\lambda_{J=1}$ = 1083.025 nm and $\lambda_{J=2}$ = 1083.034 nm  is scanned using a Toptica DFB (Distributed Feedback) pro L laser (using a LD-1083-0070-DFB-1 laser diode). The laser head is equipped with a fiber coupling unit including optical isolation in order to realize reliable single-mode operation and tuning. The line width of the laser is below 1 MHz, which is much smaller than the width of the absorption lines ($\sim$11-12 GHz) established by the strong collisional broadening at atmospheric pressure. The laser system is operated in the current controlled mode and allows linear (mode hop free) scanning over the two triplet transitions within 22 GHz. The repetition frequency of the frequency sweep is set to 1 Hz. 

\begin{figure}[!htb]
\begin{center}
\includegraphics[width=0.95\textwidth]{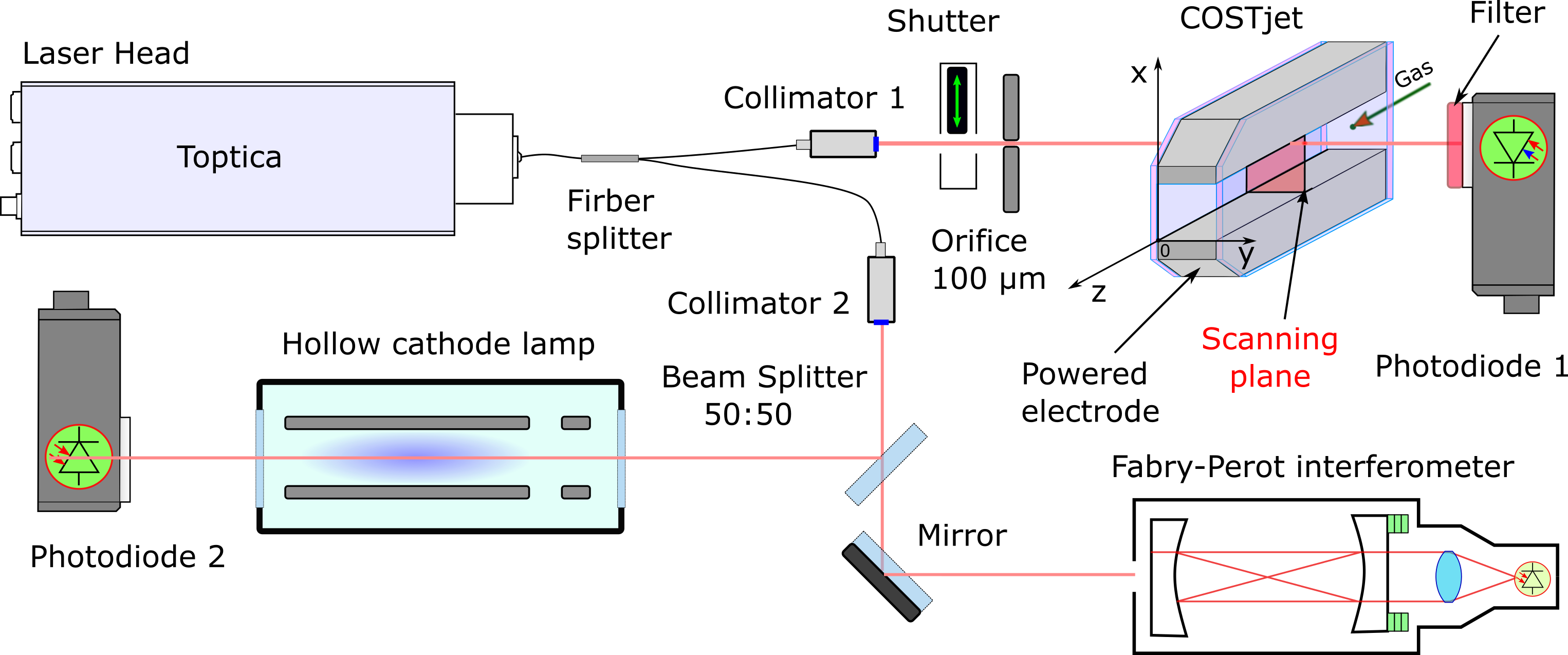}
\end{center}
\caption{Schematic view of the TDLAS set-up for the helium metastable density measurements.}
\label{fig:setupTDLAS}
\end{figure}

\noindent
In the first step of the experiment, the laser beam is split via a single-mode fiber splitter. Then, collimators form two separate beams with a diameter of 2 mm, see figure~\ref{fig:setupTDLAS}. The first beam goes through a computer-controlled shutter and then the size of the beam is reduced to 100 $\mu$m. This beam passes through the COST-jet, which is placed on a motorized stage (Standa 8MT173V-10, with a step resolution of 1.25 $\mu$m) and is detected by  "Photodiode 1" (Thorlabs DET10N2, 500-1700nm, 5 ns Rise Time). The stage is equipped  with a controller (Standa 8SMC5-USB) allowing us to set a precise position of the plasma source in $x$-direction (see figure~\ref{fig:setupTDLAS}) during the measurements. In this work, the step size is set between 15-25 $\mu$m. The photo-diode is covered by a long-pass filter (cut-on wavelength of 1000 nm) to block visible light and to improve the signal-to-noise ratio. The position $z$ (along the direction of the gas flow) is fixed at -15 mm (0 mm is the position of the jet nozzle), similar to previous measurements reported in \cite{Korolov2020, Korolov2021}. The angles at which the beam enters the jet through the quartz planes ($xz$-plane) are 90$^{\circ}$ with respect to the $x$-axis and 73$^{\circ}$ relative to the direction of the gas flow ($z$-axis). In this way unwanted Fabry-Perot resonances are avoided in the jet, that could otherwise occur, since it can act as a plane parallel optical resonator. The absorption length, $l$, is 1.05 ($\pm0.01$) mm. The detected  power of the laser beam is less than 4 $\mu$W.  By measuring the laser intensity profile along the gap, the exact electrode gap and the beam size are obtained to be 0.97 ($\pm$0.02) mm and 99 ($\pm$6) $\mu$m, respectively. 

\noindent
The second beam is split into two, equal intensity beams. One of these is guided towards a scanning Fabry-Perot interferometer with a built-in photodetector (Toptica FPI 100-1625  with 1 GHz free spectral range) to measure the line-width of the absorbing transitions. The other beam passes through a hollow cathode lamp and is detected by "Photodiode 2" (Thorlabs DET10N2), facilitating the determination of the absolute wavelength of the laser.  

\noindent
While scanning the laser wavelength over the absorption profile, the data from the two photo-detectors are recorded with a resolution of 2 MHz using another digital oscilloscope (Picoscope 6402C). Each of these signals is saved and averaged for 16-64 laser scans at every position ($x$). 
Absolute metastable densities are obtained based on the Beer-Lambert law. More detailed information about the data evaluation procedures and the accuracy of such measurements can be found elsewhere \cite{Korolov2020}.

\noindent
Day-to-day reproducibility of the measurements was in the range of 10-15\%, verifying the stability of the experimental set-up. The estimated accuracy of the measured helium metastable densities is around 10 $\%$ in addition to the background noise level that generates an error of $\pm1.5\times10^{10}$ cm$^{-3}$.

\subsection{PIC/MCC simulations}

The numerical description of the plasma is accomplished by our electrostatic "1d3v" Particle-in-Cell / Monte Carlo Collisions \cite{Birdsall1991,PIC2,PIC3} (PIC/MCC) code that has already been  used in our previous studies of atmospheric-pressure plasma jets \cite{Gibson2019, Bischoff2018}. This code implements a relatively simple model of the He--N$_2$ discharge plasma without considering an extensive set of plasmachemical reactions. The limited number of species and reactions was, however, proven to be sufficient in our previous studies for a successful reproduction of several characteristics of this plasma source (e.g., spatio-temporal distributions of the electron impact excitation rate, electron power absorption mode transitions, effects of the admixture concentration and driving voltage waveform, absolute metastable atom densities).   

The code traces electrons, He$^+$, He$_2^+$, as well as N$_2^+$ ions. The cross sections for electron impact collisions are adopted from \cite{he-cs} (for e$^-$--He atom collisions) and from \cite{n2-cs} (for e$^-$--N$_2$ collisions). We note that the latter set is based on the Siglo cross section set, available via the LxCat website \cite{siglo}. We adopt isotropic scattering for all electron-atom and electron-molecule collisions, and use, correspondingly, the elastic momentum transfer cross sections for the elastic collisions. 50\% of the e$^-$--He excitation events is assumed to lead to the formation of singlet (2$^1$S) or triplet (2$^3$S) metastable states either by direct excitation to these levels or via cascades from higher-lying states \cite{donko-nsec,Korolov2019}. As we are interested in small N$_2$ concentrations within the gas mixture, we disregard the collisions of the ionic species with the N$_2$ constituent of the mixture. For He$^+$-He elastic collisions an isotropic channel and a backward scattering channel is considered as recommended in \cite{Phelps} , while for He$_2^+$--He and N$_2^+$--He collisions we use the Langevin cross sections of the respective processes. The ions originate either from 
\begin{itemize}
    \item electron impact ionization:
\begin{eqnarray}
{\rm e}^- + {\rm He}  \rightarrow {\rm e}^- + {\rm e}^- + {\rm He}^+, \\
{\rm e}^- + {\rm N}_2  \rightarrow {\rm e}^- + {\rm e}^- + {\rm N}_2^+,
\end{eqnarray}
\item ion conversion:
\begin{equation}
{\rm He}^+ + {\rm He} + {\rm He} \rightarrow {\rm He}_2^+ + {\rm He},
\end{equation}
\item or through Penning reactions:
\begin{equation}
{\rm He^*} + {\rm N}_2 \rightarrow {\rm He} (1^1{\rm S}) + {\rm N}_2^+ + {\rm e}^-.
\end{equation}
\end{itemize}

Regarding the last two reactions, random lifetimes of He$^+$ and He$^\ast$ are computed from the reaction rates of these processes  \cite{Brok,Sakiyama} and the given reaction is executed for each simulation particle after these lifetimes elapse. 

\noindent
The stability and accuracy conditions of the PIC/MCC scheme dictate to use a very small, $\Delta t_{\rm e} = 4.5 \times 10^{-14}$s time step for the electrons under the conditions of their highly collisional transport. This ensures that the collision probability in a time step, $P = 1- \exp(-\nu \Delta t)$ ($\nu$ being the collision frequency), stays in the order of $\cong 0.1$. Ions are traced with larger time steps in order to optimise the simulation time, for He$^+$, He$_2^+$, and N$_2^+$ ions, these are $\Delta t_{\rm He^+} = 10 \, \Delta t_{\rm e}$ and $\Delta t_{\rm He_2^+} = \Delta t_{\rm N_2^+} = 100 \,\Delta t_{\rm e}$. 

\noindent
We use a buffer gas temperature of $T_{\rm g}$ = 300 K and keep the pressure at $p$ = 1 bar. The probability of the reflection of the electrons at the electrode surfaces is set to $\alpha =$ 0.6 \cite{Bischoff2018,Hemke2013,bookrefl}. The ion-induced secondary electron emission coefficients are set to $\gamma_{\rm He^+} =$ 0.3, $\gamma_{\rm He_2^+} =$ 0.2, and $\gamma_{\rm N_2^+} =$ 0.1 \cite{Bischoff2018,Korolov2019}. Further details of the discharge model and its computational implementation can be found in \cite{donko-nsec,Gibson2019, Bischoff2018}.

\section{Results}

\subsection{Dissipated power}

\begin{figure}[!htb]
\begin{center}
\includegraphics[width=0.5\textwidth]{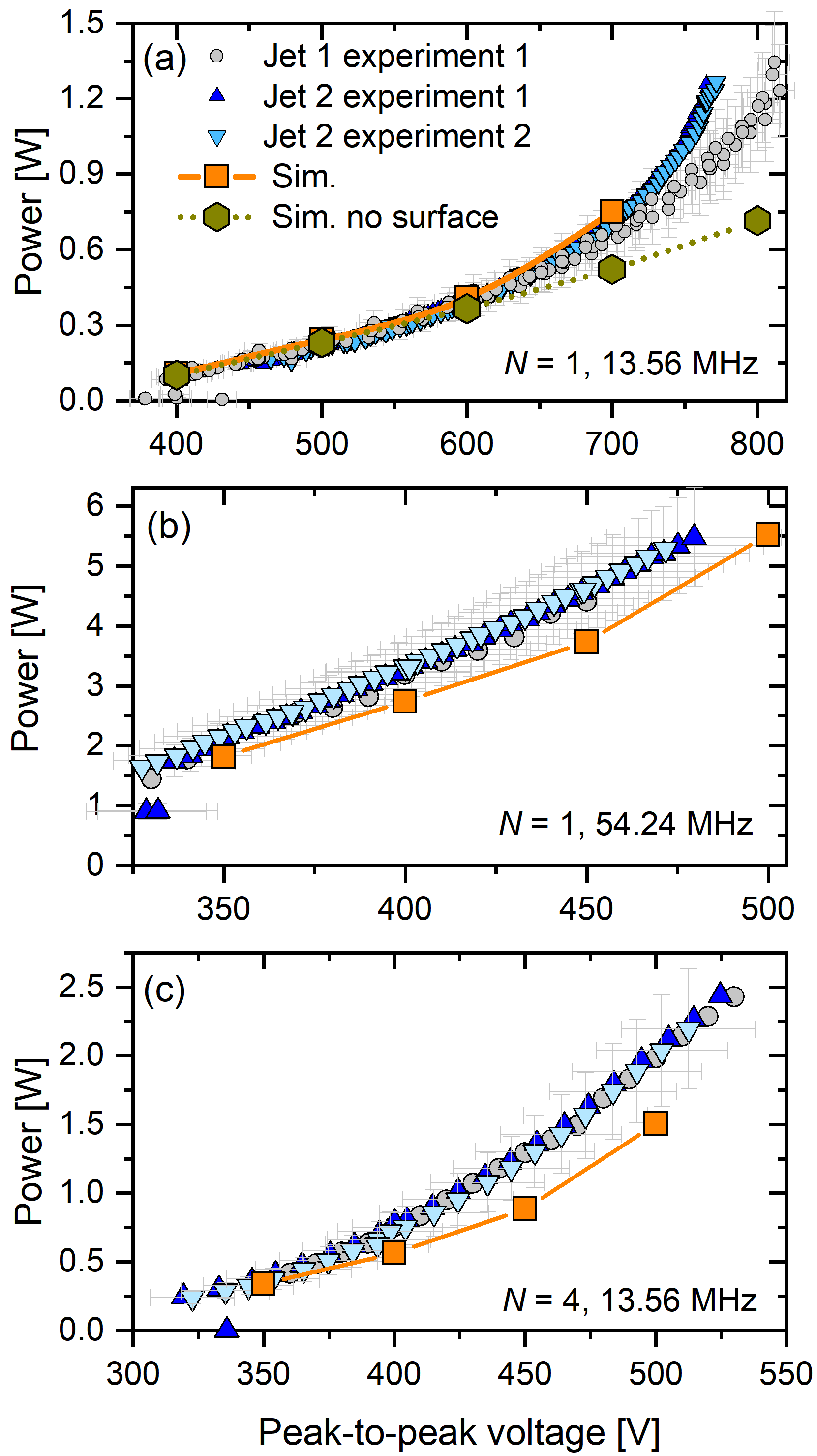}
\end{center}
\caption{Experimental and simulation data of the power dissipated in the COST-jet as a function of the peak-to-peak driving voltage: (a) single frequency - $f_0$ = 13.56 MHz, (b) single frequency - $f_0$ = 54.24 MHz, (c) ``valleys''-waveform for $N = 4$ and  $f_0$ = 13.56 MHz. The discharge gas contains 0.1 \% of N$_2$. In the experiment, two different COST-jets (Jet 1 and 2) are used and experiments are performed at different times for a given jet (experiment 1 and 2). In the simulation, the surface coefficients for ion induced secondary electron emission and electron reflection at the electrodes are switched on and off (indicated as "no surface" in the legend). }
\label{fig:powerAll}
\end{figure}

Figure~\ref{fig:powerAll} shows the measured and computationally obtained power dissipated in the plasma as a function of the applied peak-to-peak voltage for different waveforms and for a 0.1\% N$_2$ admixture to helium in the COST-jet. The data are recorded using two COST-jet devices to verify the sensitivity of the experimental results to eventual minimal differences of the plasma source. At peak-to-peak voltages of about 700 V, for the single-frequency case at $f = 13.56$ MHz (see figure~\ref{fig:powerAll} (a)), there is a clear deviation of the measured power between the two jets. "Jet 1" was previously operated for at least 800 hours, while "Jet 2"was newly built prior to the present measurements.Our measurements conducted for a period of about two months indicate that the difference mentioned above can be attributed to the properties of the two plasma sources (small manufacturing tolerances for the electrode gap, alignment, or surface properties of the electrodes) and do not originate from the instability of the experimental system. Similar observations were mentioned by Riedel et al. in \cite{Riedel2020}, where four COST-jet devices were compared at identical operational conditions. 

\noindent
For the COST-jets in the present He/N$_2$ mixture, the operating (peak-to-peak) voltages are limited to $\sim 750-800$ V. Above these voltages the discharge switches into a "constricted" mode where an unstable filament can be observed between the electrodes. This mode is characterized by high powers and it is undesired as it leads to damage of the jet via excessive gas heating. 

\noindent
In order to compare the measured power with 1D simulation results, the calculated mean power densities over one RF period are multiplied by the discharge volume of the jet ($\sim30$ mm$^3$). 
The computed power is a sum of two components that represent power absorption by electrons and ions. Both are computed as $jE$, where where $j$ is the current density of the given particle and $E$ is the electric field. The current density is obtained as $j = \pm e nu$, where $e$ is the elementary charge, $n$ is the particle density and $u$ is the mean  velocity. Despite the simplified approach, very good agreement of the experimental and computational results is generally found for all conditions studied here. For the single frequency case at $f_0 = 13.56$ MHz, the dissipated power increases linearly with the peak-to-peak voltage until around 600 V. At such low powers the discharge is mainly operated in the $\Omega$-mode, see e.g. \cite{Bischoff2018, Iza2007, Dunnbier2015}.  In this mode, a strong electric field is generated inside the bulk at the times of high current within each RF period, since the conductivity is low due to a high number of electron-neutral collisions. Thus, electron power absorption and ionization/excitation mainly occur inside the plasma bulk at the times of maximum current within the RF period.  

\noindent
At higher voltages, in Penning mixtures, e.g. He/N$_2$(O$_2$), the discharge is operated in the ``Penning"-mode \cite{Bischoff2018, Liu2021}. It is based on two main ionization/excitation pathways, which are both attributed to Penning ionization, a direct and an indirect one  \cite{Bischoff2018}. The direct channel corresponds to ionization/excitation caused by electrons directly generated by Penning ionization inside the sheaths. These electrons are then accelerated towards the bulk by the sheath electric field and can cause ionization/excitation. The second channel is indirect: positive ions, created via the Penning ionization process inside the sheaths,  are accelerated towards the adjacent electrode by the sheath electric field and, upon impact, induce the emission of secondary electrons. These electrons are also accelerated towards the bulk by the sheath electric field and can cause ionization/excitation inside the sheath. 

\noindent The simulations allow to observe the effects of the surface processes on the discharge characteristics by turning these processes (the ion-induced secondary electron emission and the elastic reflection of electrons from the electrodes) on and off. As figure~\ref{fig:powerAll}(a) shows, in the absence of surface processes the increase of the absorbed power remains linear with the peak-to-peak voltage even beyond 600 V. The observations also imply that the direct Penning ionization is not strong enough to cause the non-linear increase of the dissipated power observed experimentally at 700 - 800 V. At these conditions in the simulation, excitation and ionization occur mainly within the plasma bulk, corresponding to the $\Omega$-mode. 

\noindent The behavior of the absorbed power with the peak-to-peak value of the driving voltage shows a characteristic change when the surface processes are included in the simulation. The coefficients characterising these processes were adjusted to get good agreement between the computational results and the experimental findings for "Jet 2" at $f_0 = 13.56$ MHz and $N = 1$. The electron reflection probability at the electrodes was set this way to $\alpha =$ 0.6 and the ion-induced secondary electron emission coefficients were set to $\gamma_{\rm He^+} =$ 0.3, $\gamma_{\rm He_2^+} =$ 0.2, and $\gamma_{\rm N_2^+} =$ 0.1. If the surface effects are taken into account via these coefficients, no convergence of the simulations can be achieved at peak-to-peak voltages above 750 V, which corresponds to the "constricted"-mode in the experiment mentioned above. At lower voltages an excellent agreement is reached between the simulation results and the experimental data.

\noindent
The effect of increasing the driving frequency on the dissipated power in single frequency discharges is depicted in figure~\ref{fig:powerAll}(b) for different peak-to-peak driving voltages. 
The frequency is chosen to be  $f_0=54.24$ MHz, which corresponds to the highest harmonic used in this work to construct ``valleys''-waveforms. In contrast to $f_0 = 13.56$ MHz, much higher power densities are dissipated in the high frequency discharge at low voltages. At 54.24 MHz the discharge cannot be operated at low powers comparable to those observed in the 13.56 MHz single frequency scenario. Already near the breakdown voltage of about 325 V, more than 1.5 W power is dissipated. Similar frequency effects were previously observed in a $\mu$APPJ by Chirokov et al. ~\cite{Chirokov2009} for a frequency range of $13.56 - 40.68$ MHz. At $\gtrsim 500$ V peak-to-peak voltage a hot filamentary discharge at the tip of the jet and a very narrow and bright emission close to the electrodes are observed. Simulations indicate, that due to the increasing number of secondary electrons there is a strong avalanche formation leading to the sheath collapse/breakdown. In this regime, the power consumption is drastically increased, which causes thermal damage of the COST-jet in the experiment.

\noindent The experimental data are again well reproduced by the PIC/MCC simulations, the small quantitative differences between the computational and the experimental results might be attributed to temperature and/or multidimensional effects, which are not captured by the 1D simulations. 

\noindent
Figure~\ref{fig:powerAll} (c) compares experimental and simulation results for ``valleys''-waveforms constructed from a fundamental frequency $f_0 = 13.56$ MHz and $N = 4$ consecutive harmonics. By switching the driving voltage waveform from ``valleys'' to ``peaks'' the absolute value of the dissipated power remains the same, therefore we limit our following discussion to ``valleys''-waveforms. In the single-frequency cases, adjusting the peak-to-peak driving voltage within the limits of a stable discharge allows to change the power consumption by a factor of $\sim 6$ for $f_0 = 13.56$ MHz and $\sim 3$ for $f_0 = 54.24$ MHz. In case of the ``valleys''-waveforms this factor is around 10, i.e. a better power range control can be realized via VWT compared to single frequency operation. The range of the absorbed power is between those obtained with single frequency excitation at 13.56 MHz and 54.24 MHz. Similar to the previous cases a hot unstable filamentary discharge at the tip of the jet and a very narrow bright emission close to the powered electrode are observed, at $\gtrsim 550$ V peak-to-peak driving voltage.

\noindent
In conclusion, by changing the base frequency and applying tailored voltage waveforms the power range of stable operation of the plasma source can be significantly extended. In the following section, we will reveal how the observed power dissipation in the COST-jet under single and VWT conditions results in the formation of helium metastables.  

\subsection{Generation of Helium metastables}

\begin{figure}[h!]
\begin{center}
\includegraphics[width=0.9\textwidth]{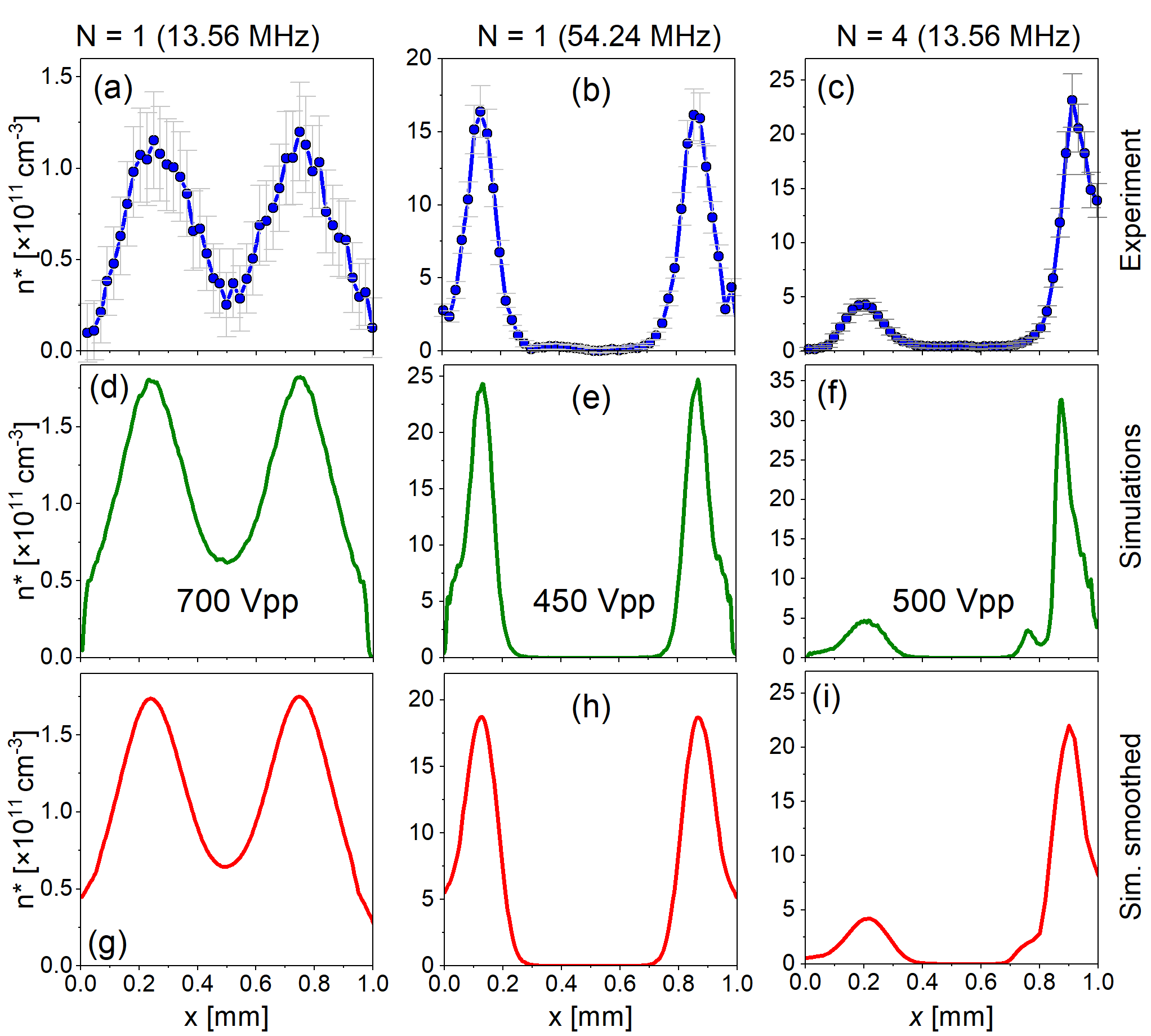}
\end{center}
\caption{(a)-(c) Measured absolute density profiles of He-I $2^3$S$_1$. (d)-(f) Time averaged helium metastables density profiles (sum of singlets and triplets) obtained from the PIC/MCC simulations. (g)-(i) Simulation data averaged over a 100 $\mu$m window to match the spatial resolution of the experiment. The powered electrode is located at $x$ = 0, while the grounded is at $x$ = 1 mm.  [N$_2$] = 0.1\%. The columns show data for different voltage waveforms: (left) single frequency - $f_0$ = 13.56 MHz and 700 V$\rm{pp}$, (middle) single frequency - $f_0$ = 54.24 MHz and 450 V$\rm{pp}$, (right) ``valleys''-waveforms for $N = 4$, $f_0$ = 13.56 MHz and 500 V$\rm{pp}$. The peak-to-peak voltages correspond to the upper limit of stable discharge operation in the simulations and in the experiments for the respective driving voltage waveform.}
\label{fig:metastables}
\end{figure}

In this section, we study the effects of different types of driving voltage waveforms on the helium metastable density profile in the COST jet and compare the energy efficiency of the generation of these species. Figure~\ref{fig:metastables} shows the measured [(a)-(c)] and computationally obtained  [(d)-(f)] spatially resolved and time averaged helium metastable densities for a 0.1 \% N$_2$ admixture to helium at the maximum achievable peak-to-peak voltages (both in the experiment and in the simulations) for the respective driving voltage waveform.  The spatial resolution of the PIC/MCC data is 2.0~$\mu$m, while in the experiment the step size of the measurements is 20~$\mu$m and the laser beam diameter is approximately 100 $\mu$m. Thus, an averaging procedure is applied to the computational results in order to make them directly comparable to the experimental data. For each measurement position $x$, that corresponds to the center of the laser beam profile, the PIC data are averaged over a region of interest from $x - 50 \, \mu$m to $x + 50 \, \mu$m around $x$ to resemble the spatial resolution of the experiment (see figures~\ref{fig:metastables}[(g)-(i)]). The measured time averaged helium metastable density profiles are in very good agreement with these computed profiles for all driving voltage waveforms after applying this averaging procedure. 

\noindent Helium atoms have two, singlet He-I 2$^1$S$_1$ and triplet He-I 2$^3$S$_1$, long-lived metastable states. In our plasma, the population of triplet metastables is expected to prevail over the singlet metastables due to an efficient conversion channel via superelastic collisions of the singlet metastables with thermal electrons ~\cite{Phelps1955, Marriott1956}:
$
\textnormal{He}(2^1\textnormal{S}_1) + \textnormal{e}^- \rightarrow  \textnormal{He}(2^3\textnormal{S}_1) +  \textnormal{e}^- + 0.79 \textnormal{ eV}.
$
As this spin changing reaction is not included in the simulations, figure~\ref{fig:metastables} [(d)-(i)] shows the sum of the densities of metastable species obtained from the simulation that is believed to be comparable to the measured He-I 2$^3$S$_1$ triplet metastable density.

\noindent
The shape of the density profile for the single frequency scenarios ($N = 1$) is spatially symmetric as a result of the symmetric spatio-temporal electron impact excitation dynamics \cite{Bischoff2018}. At high peak-to-peak driving voltages, the discharge is operated in the Penning-mode where metastables are mainly generated inside the sheaths and close to the electrodes. The main destruction mechanism of the metastables is Penning ionization of nitrogen molecules and the reaction time of this process is much shorter than the characteristic diffusion time for the metastables. Thus, the profile of the measured densities along the gap should be similar to that of the electron impact excitation rate, for details see also Korolov et al. \cite{Korolov2020}. 

\noindent
The increase of the (single) driving frequency from 13.56 MHz to 54.12 MHz leads to a higher metastable density (from $1.8\times10^{11}$ cm$^{-3}$  at 700 V$\rm{pp}$ and 13.56 MHz to $4.6\times10^{12}$ cm$^{-3}$ at 500 V$\rm{pp}$ and 54.12 MHz  ). At the same time the plasma  density also increases which leads to a shorter sheath width. Consequently, the metastables are generated closer to the electrodes. The PIC/MCC simulations show that the highest electron/ion density, which can be reached for stable operational conditions, is   $1.7\times10^{11}$ cm$^{-3}$ at $f_0 = 13.56$ MHz ($N=1$, 700 V$\rm{pp}$) and   $2\times10^{12}$ cm$^{-3}$ at $f_0 = 54.12$ MHz ($N=1$, 500 V$\rm{pp}$).

\noindent
By increasing the number of harmonics, $N$, i.e. by using Voltage Waveform Tailoring, the spatio-temporal electron impact excitation dynamics can be tailored. If ``valleys''-waveforms are used to drive the discharge, these dynamics exhibit a strong spatial and temporal asymmetry which enhanced as a function of $N$.  The electron power absorption is confined to a small region in both space and time close to the grounded electrode during the local sheath collapse, which breaks the spatial symmetry of the density profile. This effect (about which more detailed information can be found elsewhere \cite{Gibson2019, Korolov2019, Liu2021VWT}) significantly enhances the production of the helium metastables close to the grounded electrode.  Figure~\ref{fig:metastables} shows that our simulation accurately describes the COST-jet under such discharge conditions including the generation of metastables. For $f_0=13.56$ MHz ($N = 4$) and 500 V peak-to-peak driving voltage, the peak electron density is found from simulations to be $0.95\times10^{12}$ cm$^{-3}$, which is approximately two times lower than for the  single frequency case where only the highest harmonic ($f_0 = 54.12$ MHz, 500~V$\rm{pp}$) of the tailored voltage waveform is used.

\begin{figure}[!htb]
\begin{center}
\includegraphics[width=0.45\textwidth]{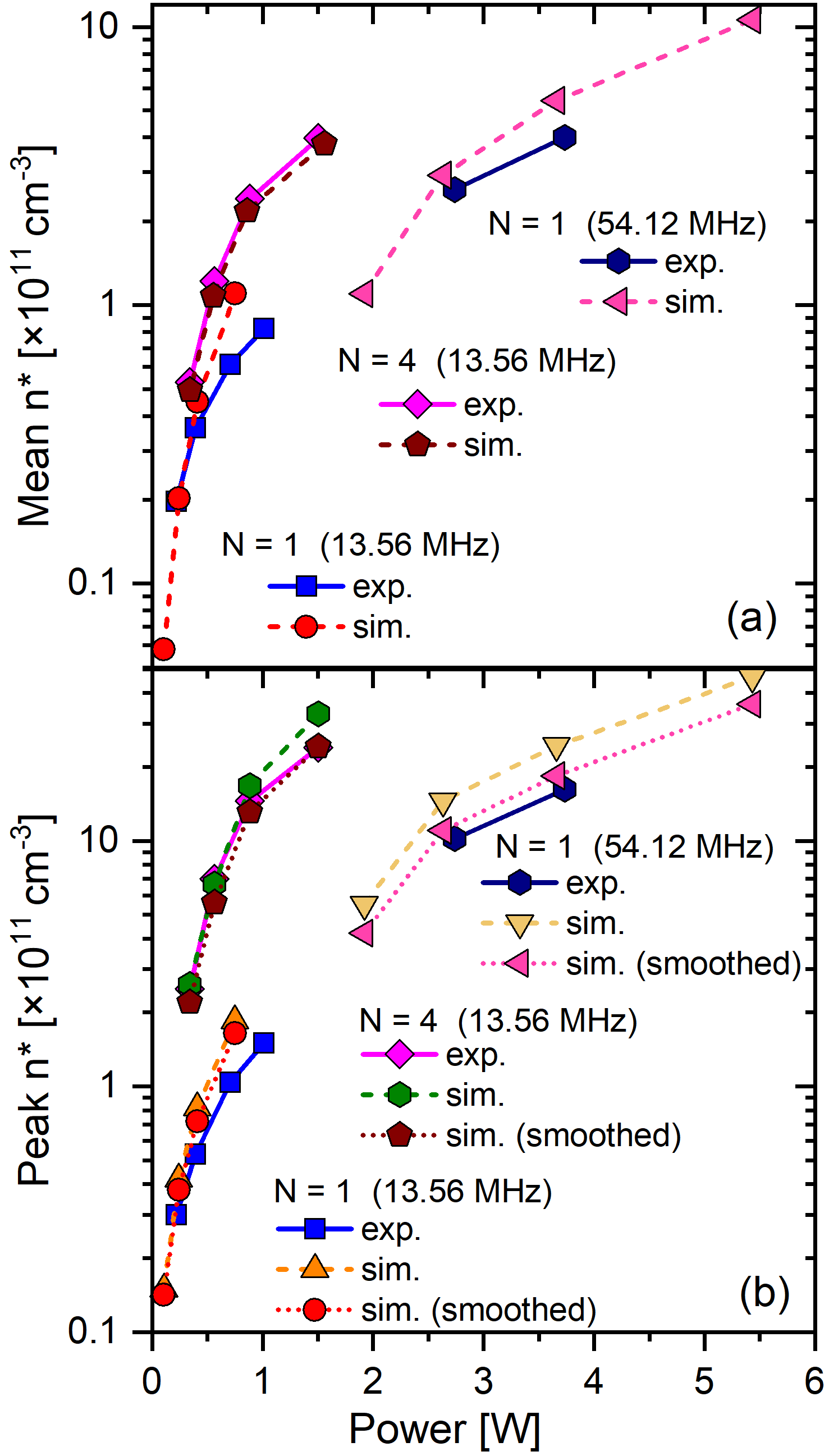}
\end{center}
\caption{Space- and time-averaged (a), and peak (b) densities of the helium metastables obtained from the experiments and the simulations as a function of the dissipated electrical power in the jet operated with different waveforms and base frequencies, $f_0$. [N$_2$] = 0.1\%. }
\label{fig:Hedenspower}
\end{figure}

\noindent
Figure~\ref{fig:Hedenspower} compares the space- and time-averaged and the peak helium metastable densities obtained from the PIC/MCC simulations and experimentally for different driving waveforms and the corresponding dissipated powers. The peak densities obtained from the simulations are calculated from the "smoothed" density profiles to allow for a direct comparison with the experimental data, see figure~\ref{fig:metastables} and the discussion above. We find excellent agreement between the experimental and the simulation data. The highest peak and mean metastable densities, which can be reached for the single-frequency case at $f_0 = 13.56$ MHz, are below $2\times10^{11}$ cm$^{-3}$.  To achieve a higher generation of helium metastables, the base frequency needs be increased or the driving voltage waveform shape must be modified. While both approaches yield higher metastable densities, the energy efficiency is significantly different for the high single frequency case ($f_0 = 54.12$ MHz) and the Voltage Waveform Tailoring case ($N = 4$). To reach high mean densities around $3\times10^{11}$ cm$^{-3}$ in the experiment, 2.8 times more dissipated power is needed for the high single frequency case compared to the "valleys"-waveform case. The difference between the dissipated power in these two scenarios is even larger for the peak densities. To reach peak metastable densities of $1\times10^{12}$ cm$^{-3}$ a factor of $\sim4$ higher dissipated power is required in the single high frequency compared to the VWT case. This observation allows us to conclude that VWT is a more energy efficient way to control and generate metastable densities. The energy efficiency of VWT to generate high metastable densities is higher by a factor of 3 - 4 compared to single frequency operation. A detailed description and understanding of this effect is provided in the next section. We note that, according to figure \ref{fig:Hedenspower}, an extrapolation of the 13.56~MHz single frequency data to higher dissipated powers approximately yields the 54.12 MHz single frequency results, i.e. all single frequency results for the helium metastable density as a function of the dissipated power lie approximately on a single curve. The same is expected for the VWT results, i.e. increasing the fundamental driving frequency is expected to result in higher dissipated powers that yield higher metastable densities according to an extrapolation of the corresponding curve in figure \ref{fig:Hedenspower}. This, however, could not be tested in the frame of this work, but remains a topic for future investigations.  

\subsection{Power dissipation dynamics in the COST-jet}

\begin{figure}[!htb]
\begin{center}
\includegraphics[width=0.45\textwidth]{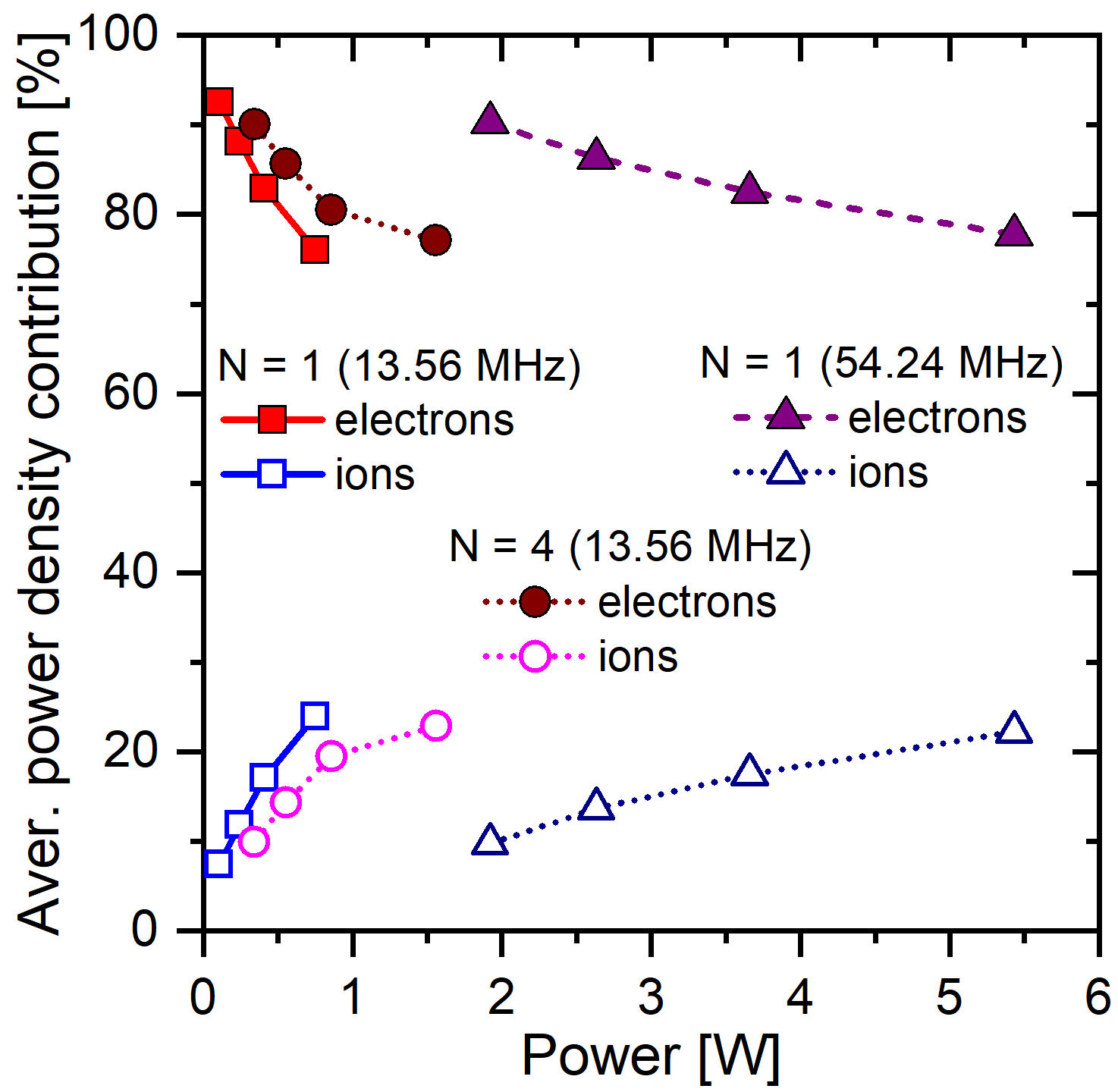}
\end{center}
\caption{Fraction of the total power dissipated to electrons and ions in the COST-jet as a function of the total dissipated power. These data are extracted from the PIC/MCC simulations of the jet operated at different waveforms and base frequencies, $f_0$. [N$_2$] = 0.1\%. }
\label{fig:powerdens}
\end{figure}

\noindent
In this section, we investigate the effects of different driving voltage waveforms on the electron and ion power absorption dynamics to explain why VWT allows to generate helium metastables in a more energy efficient way as compared to classical single frequency operations of $\mu$APPJs. Based on PIC/MCC simulation data, figure~\ref{fig:powerdens} shows the fraction of the total power dissipated to the electrons and to the ions in the COST-jet as a function of the total dissipated power.  Our findings are that
\begin{itemize}
    \item in contrast to low pressure capacitively coupled plasmas, where a large fraction of the total power is dissipated to the ions \cite{Popov1985, Lafleur2012, Rauf2009, Wilczek2020}, at atmospheric pressure the power is mostly dissipated to electrons under the conditions studied here;
    \item the power dissipated to the ions is, however, not negligibly small and varies from 7.5\% up to 25\% for the conditions studied. There are numerous  fluid, global or electrical approaches of atmospheric pressure radio-frequency capacities discharges \cite{Lazzaroni2012, Lazzaroni2012a, Waskoenig2010, Golda2019, Shi2005}. These models are usually limited to low powers, where the discharge is operated in the $\Omega$-mode, or/and they neglect mobility-driven ion fluxes towards the electrodes. According to our results, the simplification that all the input power is absorbed by the electrons only may lead to significant deviations of the obtained plasma parameters from the measured ones. Thus, it is recommended to always consider in the models the power absorbed by ions.
\end{itemize}

\begin{figure}[!htb]
\begin{center}
\includegraphics[width=1.0\textwidth]{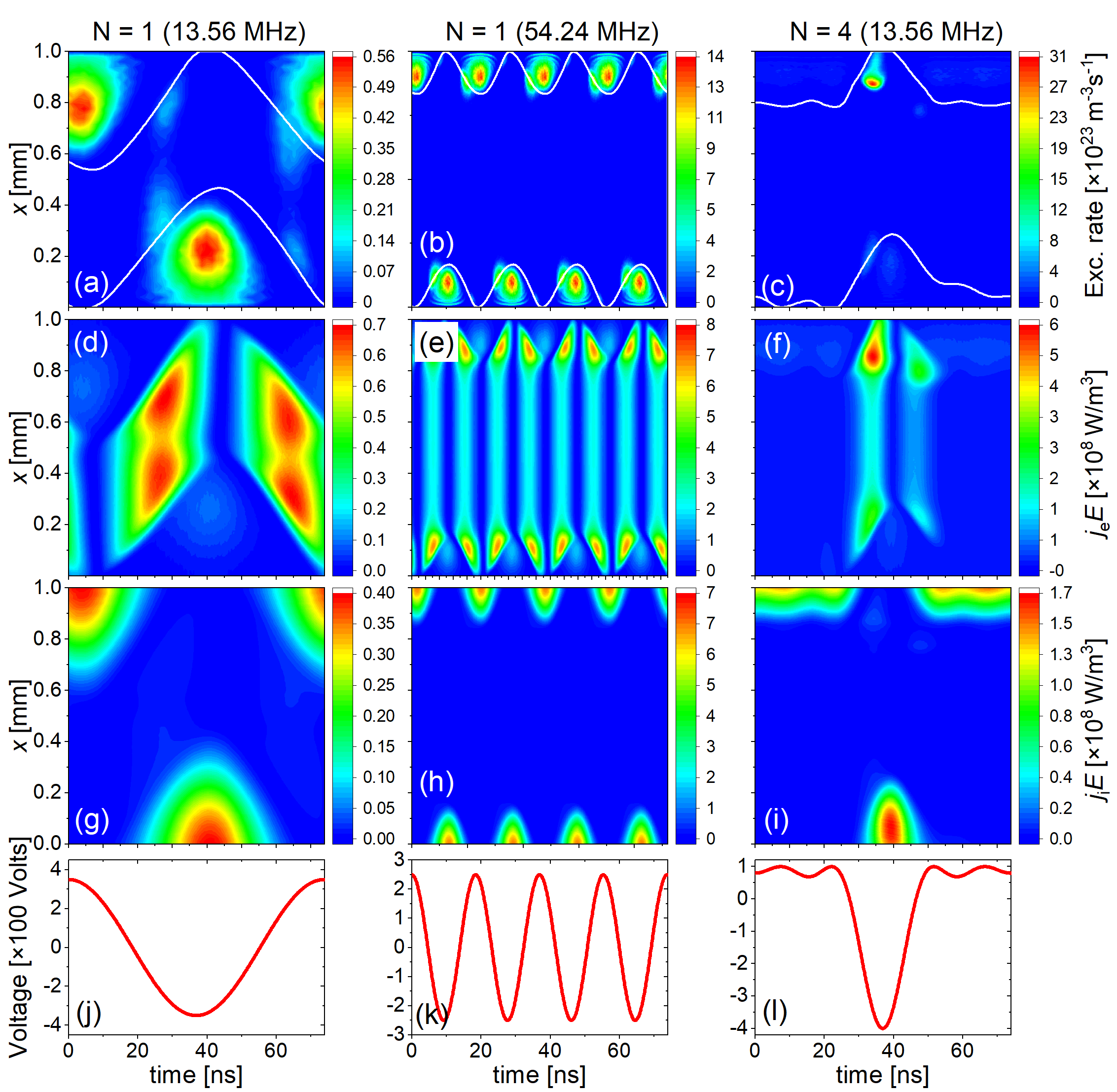}
\end{center}
\caption{ Spatio-temporal distributions of the computed electron impact excitation rate from the ground state into the He-I (3s)$^3$S$_1$-state (a)-(c), power densities absorbed by the electrons (d)-(f) and the ions (g)-(i). Time covers one period of the lowest driving frequency of $f_0$ = 13.56 MHz. The data are plotted for different driving voltage waveforms (j)-(l): (first column) single frequency - $f_0$ = 13.56 MHz and 700 V$\rm{pp}$, (second column) single frequency - $f_0$ = 54.24 MHz and 500 V$\rm{pp}$ and (third column) ``valleys''-waveforms for $N = 4$, $f_0$ = 13.56 MHz and 500 V$\rm{pp}$. (j)-(l).  The white lines on the maps mark the positions of the sheath edges determined according to \cite{Brinkmann2007}. [N$_2$] = 0.1\%. The powered electrode is located at $x = 0$, while the grounded electrode is at $x = 1$ mm.}
\label{fig:xtpower}
\end{figure}

\noindent
In our previous studies, the He metastable source was shown to be directly correlated with the spatial distribution of the  electron  impact excitation  rate  from  the  helium  ground  state  into  other  high-lying  excited  He  levels \cite{Korolov2020}.  One of these is the He-I (3s)$^3$S$_1$ level that has an excitation energy of 22.7 eV and emits radiation with a wavelength of 706.5 nm which is often chosen in experiments to characterize the plasma by Phase Resolved Optical Emission Spectroscopy (PROES) \cite{Bischoff2018, Korolov2019}. Figure~\ref{fig:xtpower}(a)-(c) shows spatio-temporal distributions of the computed electron impact excitation rate from the ground state into the He-I (3s)$^3$S$_1$-state for different waveforms. The peak-to-peak driving voltages are chosen to be close to the maximum values that allow generating stable discharges in the experiments and in the simulations. 

\noindent
According to the findings of the previous section, the single frequency scenario ($N$ = 1) is characterized by a spatially symmetric pattern of the metastable densities. This is caused by the symmetric spatio-temporal electron impact excitation dynamics shown in figure~\ref{fig:xtpower}(a) and (b):  Two excitation maxima with the same intensity are observed inside the expanded sheath at each electrode at an equal distance from the adjacent electrode within each RF period. Under these conditions, the discharges operate in the Penning-mode. With increasing frequency, the position of the excitation maxima shift towards the electrodes due to the increased plasma density and the decreased sheath width. The positions of these peaks, as expected, are at similar distances from the adjacent electrode as the positions of the maxima of the metastable atom density depicted in figure~\ref{fig:metastables}. By increasing the number of consecutive harmonics to $N = 4$ to generate the ``valleys" driving voltage waveform, the dynamics of the boundary sheaths within the fundamental RF period can be customized and the symmetry of the spatio-temporal excitation dynamics can be broken in space and time \cite{Gibson2019,Korolov2019, Korolov2020, Liu2021}. Such a waveform, depicted in figure~\ref{fig:xtpower}(l), induces a short and fast sheath collapse at the grounded electrode, while at the powered electrode the sheath collapses for a long fraction of the fundamental RF period  (figure~\ref{fig:powerAll}(c)). In order to compensate the continuous positive ion flux to the grounded electrode within one fundamental RF period, a high electron current has to be driven to this electrode during the short sheath collapse phase. To induce such an electron current a strong electric field is generated, while the sheath is collapsed \cite{SchulzeChargeDyn}. This field  can accelerate electrons to energies high enough to cause strong excitation/ionization including metastable generation in the vicinity of the grounded electrode. Therefore, in the experiment and in the simulation, see figure~\ref{fig:metastables}[(c),(f) and (i)], the peak metastable density is found to be strongly enhanced  close to this electrode for this tailored driving voltage waveform.

\noindent
The first row of figure~\ref{fig:xtpower} shows the spatio-temporal distribution of the electron impact excitation rate from the ground state into the He-I (3s)$^3$S$_1$-state, which is highly correlated with the generation of helium metastables, while the second and the third rows show the power densities dissipated to electrons and ions, respectively. Analyzing these plots allows to understand why VWT, i.e. the use of valleys driving voltage waveforms, is a more energy efficient way to generate the metastables as compared to single frequency operation. Based on a comparison of the first and the second rows of figure~\ref{fig:xtpower}, we can state that in both single-frequency cases only a small fraction of the total power dissipated to electrons results in the generation of helium metastables, while a much larger fraction is used for this in the VWT scenario. Generally, most power is dissipated to electrons inside the plasma bulk, where the electron density is high, so that the power dissipated per electron can be low and insufficient to accelerate electrons above the threshold energy to generate helium metastables by electron impact (ca. 19.8 eV) despite the high total electron power absorption in the bulk. This is true for both single frequency cases as documented by the fact that there is almost no excitation in the bulk, although most power is dissipated to electrons there. In fact, most excitation and helium metastable generation occurs inside the sheaths in the single frequency scenarios, although only little power is dissipated to electrons inside the sheaths. This is caused by the fact that the electron density inside the sheaths is extremely low so that the energy gain per electron is high, although the local total electron power absorption is low. Such electrons inside the sheaths are either directly generated by Penning ionization or indirectly by secondary electron emission from the electrodes due to the impact of positive ions generated by Penning ionization. Such electrons are accelerated by the high sheath electric field into the bulk and gain enough energy to cause excitation and generation of metastables. 

\noindent In principle, these same mechanisms are present in the low and high single frequency cases. The only difference is the fact that the RF period is shorter in the 54.12 MHz compared to the 13.56 MHz case and, thus, the sheaths oscillate more frequently per unit time in the high frequency case. Thus, more power is dissipated in the high frequency case and higher metastable densities can be produced, but the energy efficiency of this process is not improved (see figure \ref{fig:Hedenspower}). Overall, in terms of helium metastable generation, the electron power absorption is inefficient under single frequency operation as most power is dissipated to electrons in the bulk in a way that each electron gains little energy, which is subsequently lost by electron-neutral collisions and is too low to generate metastables. 

\noindent This is very different in case of VWT, i.e. valleys driving voltage waveforms, as shown in the third column of figure \ref{fig:xtpower}. Due to the specific sheath dynamics induced by this particular waveform most power is now dissipated to electrons within a narrow domain of space and time within the fundamental RF period. Thus, a large fraction of the total power is now dissipated to a small group of electrons located at the corresponding spatial position. This position is located outside the sheath, where the electron density is relatively high. Correspondingly, a high number of electrons gains enough energy to generate helium metastables. This explains why a given power dissipated to electrons results in higher metastable densities in case of the valleys driving voltage waveform as compared to single frequency operation. Thus, the energy efficiency of helium metastable generation is higher for VWT compared to the classical single frequency operation of $\mu$APPJs.

\begin{figure}[!htb]
\begin{center}
\includegraphics[width=0.45\textwidth]{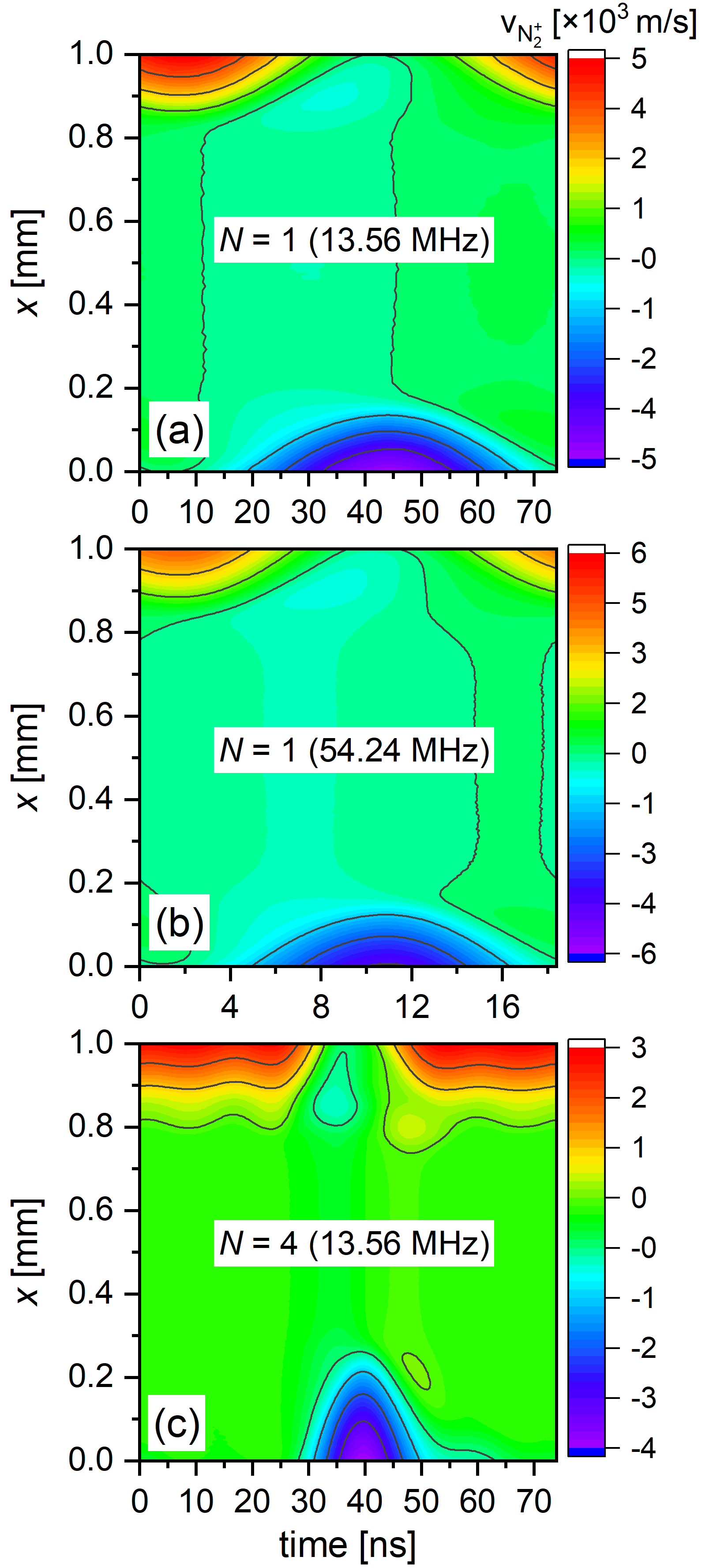}
\end{center}
\caption{Spatio-temporal distribution of the mean velocity of the N$_2^+$ ions obtained from the simulations for different driving voltage waveforms: (a) single frequency - $f_0$ = 13.56 MHz and 700 V$\rm{pp}$, (b) single frequency - $f_0$ = 54.24 MHz and 500 V$\rm{pp}$ (c) ``valleys''-waveforms for $N = 4$, $f_0$ = 13.56 MHz and 500 V$\rm{pp}$, [N$_2$] = 0.1\%}
\label{fig:xtvel}
\end{figure}

\noindent
Figures~\ref{fig:powerdens} and ~\ref{fig:xtpower} (third row) show that the power dissipated to ions is not negligibly small. The electric field inside the sheaths is much stronger than the electric field inside the bulk for all presented cases. This strong sheath electric field accelerates ions towards the electrodes, as shown in figure~\ref{fig:xtvel}, which leads to an enhancement of the ion current density inside the sheath and, thus, to an increase of the dissipated power. With increasing applied voltage the ion density and the electric field inside the sheath increase leading to an enhancement of the total power dissipated to ions, see figure~\ref{fig:powerdens}. In low-pressure RF discharged the ions are usually assumed not to follow the RF electric field due to their high mass. Instead, they only react to the time averaged electric field. At atmospheric pressure, this assumption is not correct, since the ion-neutral collision frequency is high, i.e. $\sim 1.4 \times 10^{10}$ s$^{-1}$ (according to an estimation based on the Langevin cross-section~\cite{Gioumousis1958}), which is much higher than the driving frequencies used in this work. The average time between such collisions is around 74 picoseconds and the ion energy relaxation time is assumed to be around 1 ns \cite{Winkler2002}. Thus, the ions can almost instantaneously follow the change of the electric field. The effect can be clearly seen in figure~\ref{fig:xtvel}, where the spatio-temporal velocity distribution of the dominant ion, N$_2^+$, obtained from the simulations is plotted for the different driving voltage waveforms. 

\noindent
Generally, the dynamics of the power dissipation to electrons and ions is important and should be taken into account in models used to describe such plasmas. Models that use only the time averaged dissipated power as an input parameter, according to figure~\ref{fig:xtpower}, need to take the time dependence of the power dissipation into account. For instance, in the case of the single-frequency waveforms, the difference between the mean and the instantaneous peak power values within one RF period is less than a factor of two, while for the ``valleys''-waveform used in this paper this factor is more than 4. Therefore, using time-averaged power values as an input for, e.g., global models \cite{Park2008, Liu2010, Hurlbatt2017}, can lead to serious miscalculations of the plasma parameters. This is distinctly visible in figure~\ref{fig:Hedenspower}: For the same dissipated power of approximately 2 W, the mean value of the helium metastable densities for single frequency operation is a factor of three lower than for the ``valleys'' waveform. 

\section{Conclusions}

The effects of various driving voltage waveforms on the energy efficiency of the generation of helium metastables in capacitively coupled atmospheric pressure microplasma jets were studied both experimentally and via PIC/MCC simulations. In the experiment, a method to determine the power dissipated to the plasma in the presence of multiple driving frequencies (Voltage Waveform Tailoring (VWT)) was developed and applied. Absolute values of the density of He-I 2$^3$S$_1$ metastables were measured spatially resolved in the COST-jet by tunable diode laser absorption spectroscopy for single (13.56 MHz and 54.12 MHz) and multi-frequency operating conditions. The measurements were performed at a fixed admixture of nitrogen to helium of 0.1\% and different peak-to-peak driving voltages. The experimental results were found to be in very good quantitative agreement with those of the PIC/MCC simulations. 

\noindent
Based on these measurements and simulations, a detailed fundamental analysis of the power dissipation to charged particles (electrons and ions) was performed at different single driving frequencies as well as valley-shaped tailored voltage waveforms. VWT was found to yield a significantly better energy efficiency of helium metastable generation compared to single frequency operation, since it allows to tailor the electron power absorption dynamics in a way that confines a large fraction of the total power dissipated to electrons to a narrow spatial domain between the electrodes and to a particular time within the fundamental RF period. In this way the energy gain per electron within this spatio-temporal region of interest is increased so that metastables are created efficiently by electron impact excitation. This is markedly different compared to the standard single frequency scenarios, where a given total power dissipation to electrons is distributed over larger regions in space and time so that the energy gain per electron is too low to generate helium metastables efficiently. This shows that VWT is not only able to provide a unique way to control neutral species densities, but also to generate them in a more energy efficient way. 

\noindent
Excited species such as helium metastables play an important role for many applications,  e.g. plasma medicine, surface modification, etc. Therefore, VWT provides new concepts to optimize such technological applications in addition to its fundamental relevance. Clearly, the potential of VWT to control and energy efficiently generate other species should be investigated. Due to other energy thresholds, different waveform shapes might be more appropriate for such purposes. 

\noindent
Finally, we demonstrated that the spatio-temporally resolved power dissipation dynamics should be considered in models to describe the generation of selected neutral species correctly rather than taking into account only a space and time averaged value of the power dissipation. This is verified by the fact that the same averaged power density was demonstrated to result in significantly different helium metastable density in case of different driving voltage waveforms due to the different spatio-temporal power absorption dynamics of charged particles. This, in fact, is the reason for the superior energy efficiency of VWT observed in this work and could be completely missed in models that only use averaged power values as an input.

\ack This work is supported by the DFG via SFB 1316 (projects A4), by the Hungarian Office for Research, Development, and Innovation NKFIH grant K-134462, and by the National Natural Science Foundation of China (Grant No. 12020101005).

\end{document}